\documentclass[aps,prd,twocolumn,groupedaddress,nofootinbib]{revtex4}  

\usepackage{graphicx}  
\usepackage{dcolumn}   
\usepackage{bm}        
\usepackage{amssymb,amsfonts,amsmath,physics}   
\usepackage{cancel}
\usepackage{mathtools}

\usepackage{tikz-feynman}
\tikzfeynmanset{compat=1.1.0}
\usetikzlibrary{arrows}
\tikzset{	
	vertex/.style={circle,draw, minimum size=1.5em},	
	edge/.style={->,> = latex'}	
}
\usepackage[bookmarks, breaklinks, colorlinks,urlcolor=blue, citecolor=red, 
linkcolor=blue]{hyperref}
\usepackage[normalem]{ulem}

\newcommand{\be}{\begin{eqnarray*}}
	\newcommand{\ee}{\end{eqnarray*}}

\newcommand{\bee}{\begin{eqnarray}}
	\newcommand{\eee}{\end{eqnarray}}
\newcommand{\beeq}{\begin{equation}}
	\newcommand{\eeq}{\end{equation}}

\usepackage{xspace}

\newcommand{\ba}{\begin{array}}
	\newcommand{\ea}{\end{array}}
\newcommand{\bd}{\begin{displaymath}}
	\newcommand{\ed}{\end{displaymath}}
\newcommand{\besub}{\begin{subequations}}
	\newcommand{\eesub}{\end{subequations}}
\newcommand{\bea}{\begin{eqnarray}}
	\newcommand{\eea}{\end{eqnarray}}

\def\q2 {q^2}

\usepackage{tikz}
\usetikzlibrary{arrows,shapes}
\usetikzlibrary{trees}
\usetikzlibrary{matrix,arrows} 				
\usetikzlibrary{positioning}				
\usetikzlibrary{calc,through}				
\usetikzlibrary{decorations.pathreplacing}  
\usepackage{pgffor}							

\usetikzlibrary{decorations.pathmorphing}	
\usetikzlibrary{decorations.markings}
\tikzset{
	vector/.style={decorate, decoration={snake}, draw},
	provector/.style={decorate, decoration={snake,amplitude=2.5pt}, draw},
	antivector/.style={decorate, decoration={snake,amplitude=-2.5pt}, draw},
	fermion/.style={draw=black, postaction={decorate},
		decoration={markings,mark=at position .55 with {\arrow[draw=black]{>}}}},
	fermionbar/.style={draw=black, postaction={decorate},
		decoration={markings,mark=at position .55 with {\arrow[draw=black]{<}}}},
	fermionnoarrow/.style={draw=black},
	gluon/.style={decorate, draw=black,
		decoration={coil,amplitude=4pt, segment length=5pt}},
	scalar/.style={dashed,draw=black, postaction={decorate},
		decoration={markings,mark=at position .55 with {\arrow[draw=black]{>}}}},
	scalarbar/.style={dashed,draw=black, postaction={decorate},
		decoration={markings,mark=at position .55 with {\arrow[draw=black]{<}}}},
	scalarnoarrow/.style={dashed,draw=black},
	electron/.style={draw=black, postaction={decorate},
		decoration={markings,mark=at position .55 with {\arrow[draw=black]{>}}}},
	bigvector/.style={decorate, decoration={snake,amplitude=4pt}, draw},
}

\tikzstyle{block} = [draw, rectangle, 
minimum height=3em, minimum width=6em]

\begin{document}
\title{Majorons Revisited: light dark matter as FIMP}

\author{Soumen Kumar Manna}
\email{soume172121044@iitg.ac.in}
\affiliation{Department of Physics, Indian Institute of Technology Guwahati, Assam-781039, India}

\author{Arunansu Sil}
\email{asil@iitg.ac.in}
\affiliation{Department of Physics, Indian Institute of Technology Guwahati, Assam-781039, India}

\begin{abstract}
	We show that Majoron, the pseudo-Nambu-Goldstone boson resulting from the spontaneous breaking of global lepton number symmetry, can present itself as a viable freeze-in type of dark matter in a mass range $\mathcal{O}$(keV-GeV), thanks to the explicit higher dimensional Lepton number breaking operator. Interestingly, the proposal is restricted within the simplest extension of the Standard Model with two singlet right-handed neutrinos and a singlet scalar so to address light neutrino mass and spontaneous breaking of lepton number symmetry respectively.  The desired amount of Majoron production takes place from the annihilations of right-handed neutrinos indicating an intriguing connection between neutrino physics and dark matter. 
\end{abstract}
\maketitle
\section{Introduction}

Despite the clear indication that neutrinos do have mass which is reminiscent of the first established departure from the Standard Model (SM) of particle physics, it is still unclear whether they are of Dirac or Majorana type in nature. The latter possibility is related to the lepton number violation (while the first one conserves it) by heavy right-handed neutrino (RHN) mass(es) and is capable of explaining the smallness associated to the neutrino mass. Interesting consequences would follow, if we promote this lepton number symmetry (LNS) to a global one, say $U(1)_{L}$, and consider it to be broken spontaneously so as to make RHNs massive. In this case, a massless Nambu-Goldstone boson (NGB) called Majoron results \cite{Chikashige:1980qk, Chikashige:1980ui, Gelmini:1980re}. While the mass of a Majoron is related to an explicit (soft) breaking of the LNS, analogous to the quark mass in the case of pion, such a particle carrying very suppressed (by the scale of symmetry breaking) interactions, in general, renders itself as a promising dark matter (DM) candidate \cite{Rothstein:1992rh,Berezinsky:1993fm,Bazzocchi:2008fh,Gu:2010ys, Frigerio:2011in,Shakya:2018qzg,Lattanzi:2013uza, Queiroz:2014yna}. 

Previous studies with such Majoron field ($\chi$) as a freeze-out kind of DM have mostly been engineered through SM Higgs $(H)$ portal interaction of the form $\lambda_{\chi H}\chi^2 H^\dagger H$, an explicit LNS breaking term, which helps in realizing the DM relic satisfaction \cite{Frigerio:2011in,Queiroz:2014yna}. Such an interaction is suggestive 
of the scalar singlet DM scenario \cite{Cline:2013gha} and obviously, there exists a one-to-one correspondence between the mass of the DM ($m_\chi$) and the LNS breaking parameter 
$\lambda_{\chi H}$. However, contrary to the natural expectation that an explicit symmetry breaking parameter should be sufficiently small in t'Hooft's sense \cite{tHooft:1979rat}, it is observed that $\lambda_{\chi H}$ needs to be large enough ($\mathcal{O}(0.1)$ or so) in order to satisfy the correct relic density. Secondly, with the XENON-1T results, the entire parameter space of $m_\chi$ ranging below TeV (except the Higgs resonance region) is basically ruled out in this minimal Majoron scenario \cite{Queiroz:2014yna,Cline:2013gha}. The Majoron also 
possesses a derivative coupling with its CP-even partner, similar to other pseudo-Nambu Goldstone Bosons (pNGB), through which it can annihilate into SM particles via $s$-channel mediated process by the CP-even scalar leading to a possible relic satisfaction. In case of pNGB as DM (not a Majoron), such a scenario can evade the direct detection experiment bounds \cite{Gross:2017dan, Azevedo:2018exj, Ishiwata:2018sdi, Huitu:2018gbc, Arina:2019tib} as a result of the small momentum transfer of DM through the derivative coupling causing suppression in elastic scattering amplitude with nuclei. However, in case of Majoron as DM, it fails to explain the stability criteria as Majorons would decay to light neutrinos ($\nu$) via 
active sterile mixing connected to the neutrino mass generation in type-I seesaw \cite{Minkowski:1977sc,Mohapatra:1979ia,Schechter:1980gr}. 

On the other hand, Majorons may also emerge as Feebly interacting massive particle (FIMP) DM \cite{Hall:2009bx} where it is produced in the early Universe from the decay of some particle, a natural choice of which is the RHN ($N$). As the Majoron interaction strength with the SM particles is expected to be suppressed by the $U(1)_{L}$ breaking scale (the seesaw scale) \cite{Escudero:2021rfi}, treating Majoron as FIMP type DM remains as a viable option \cite{Frigerio:2011in,Garcia-Cely:2017oco,Brune:2018sab}. However, from the previous studies, it becomes apparent that with the tree-level coupling between the RHNs and the scalar singlet (the one responsible for breaking the LNS), sufficient production of Majorons cannot take place from the decay of RHNs \cite{Abe:2020dut}. This is primarily because the decay rate ($N \rightarrow \chi \nu$) is proportional to the tiny light neutrino mass on top of the usual suppression by LNS breaking scale. However, the situation changes if a tiny LNS breaking Higgs-portal coupling $\lambda_{\chi H}$ is introduced. The relic satisfaction\footnote{Direct detection bound is automatically evaded due to the presence of such a small portal coupling, contrary to the freeze out case, involved.} of Majorons produced from the decay of the SM Higgs boson $h \to \chi \chi$ requires the portal coupling $\lambda_{\chi H} \sim O(10^{-10})$ which in turn fixes the Majoron mass $m_\chi$ having a unique value $\sim$ 3 MeV \cite{Frigerio:2011in,Garcia-Cely:2017oco,Brune:2018sab}. Though such small portal coupling obeys the naturalness criteria, it suffers from a fine-tuned situation in terms of Majoron mass.  Another possibility is to realize non-thermal Majoron production from the annihilation of SM Higgs via UV freeze-in framework \cite{Abe:2020dut} where the Majoron mass is found to be relatively heavy (contrary to the light pNGB boson in UV freeze-in scenario \cite{Abe:2020ldj}). 

In this work, with the aim of broadening the Majoron mass range (toward the lighter side) while the minimality and naturalness are retained, we propose a new production mechanism for Majoron as a FIMP-type DM.  Instead of introducing any Higgs-portal coupling with Majorons, we introduce here a dimension-5 LNS breaking operator. Being suppressed by the cut-off scale, it can be regarded as a soft breaking of LNS which takes care of the production of Majorons via RHN annihilations. Also, we keep the framework minimal in the sense that no additional fields other than the two SM singlet right-handed neutrinos (related to neutrino mass generation) and a LNS breaking (SM singlet) red complex scalar field, are required.  We primarily focus on the lighter side of Majoron mass $[\mathcal{O}(\rm{keV})-\mathcal{O}(\rm{GeV})]$, which remains attractive as this mass regime can be experimentally probed by several direct and indirect searches. For example, Majorons of MeV scale and beyond can be potentially detectable via mono-energetic neutrino flux in experiments like Borexino\cite{Borexino:2010zht}, KamLAND\cite{KamLAND:2011bnd}, and Super-Kamiokande (SK)\cite{Super-Kamiokande:2002exp}. Also, sensitive bounds from various $\gamma$-ray observations such as INTEGRAL \cite{Boyarsky:2007ge}, COMPTEL/EGRET \cite{Yuksel:2007dr}, Fermi-LAT \cite{Fermi-LAT:2015kyq} etc. are applicable in this mass regime.

\section{Majoron as Dark matter } \label{section2}

\begingroup
\setlength{\tabcolsep}{10pt} 
\renewcommand{\arraystretch}{1.5}

As we build our framework extending the original Majoron model where a Majoron can be considered as a dark matter \cite{Berezinsky:1993fm,Rothstein:1992rh}, a brief discussion on the basic structure of it and related limitations are relevant to discuss first. The SM is extended by including two singlet right-handed neutrinos ($\mathcal{N}_{R_{i=1,2}}$) and a singlet complex scalar field ($\Phi$) such that a global lepton number symmetry $U(1)_{L}$ prevails. As a result, a lepton number of $L = -2$ units is assigned to $\Phi$ while the SM lepton doublets (${L}_\alpha$) as well as the RHNs carry a lepton number $L=1$. The renormalizable part of the Lagrangian (in the charged lepton diagonal basis) involving neutrinos is given by
\beeq
-\mathcal{L}\supset  \frac{f_{i}}{2}\Phi \overline{\mathcal{N}^c_{R_i}} \mathcal{N}_{R_i}+y^\nu_{\alpha i}~\overline{L}_\alpha \tilde{H}\mathcal{N}_{R_i} +h.c.,  
\label{lag}
\eeq
where, $H$ is the SM Higgs doublet and $y^\nu_{\alpha i}$ is the neutrino Yukawa coupling with $\alpha=e,\mu,\tau$ and $i =1,2$. The Majorana masses of RHNs are generated after the $U(1)_L$ symmetry breaking via $\Phi$ vacuum expectation value (vev) $v_{\phi}$. Without loss of generality, we consider the generated RHN mass matrix $M_R = f v_{\phi}/\sqrt{2}$ to be diagonal with $f ={\rm{diag}} (f_1, f_2)$. Once the electroweak (EW) symmetry is broken by the Higgs vev $v$ as $\langle H \rangle = (0, v/{}\sqrt{2})^T$, active neutrino masses are generated through this minimal type-I seesaw mechanism \cite{Ibarra:2003up} {$m_\nu=m_D M_R^{-1} m_D^T$, where $(m_D)_{\alpha i}=y^\nu_{\alpha i}v/\sqrt{2}$ $(v=246~ \textrm{GeV})$}. 

The $U(1)_L$ symmetric scalar potential involving $H$ and $\Phi$ is given by,
\beeq V(H,\Phi)=V_H - \frac{{\mu_\Phi^2}}{2}|\Phi|^2+ \frac{\lambda_\Phi}{2} |\Phi|^4+\lambda_{H\Phi}|H|^2|\Phi|^2,
\label{pot} 
 \eeq
where {$V_H = -\mu_H^2 H^\dagger H+\lambda_H (H^\dagger H)^2$}  is the usual SM Higgs potential and $\lambda_{H\Phi}$ is the Higgs portal coupling of $\Phi$. The stability of the potential is guaranteed by: $\lambda_{\Phi},\lambda_H ~{\rm{and}}~ \lambda_{H\Phi}+\sqrt{2\lambda_H \lambda_\Phi}\geq 0.$

Once the $U(1)_L$ global symmetry is broken spontaneously, $\Phi = (v_\phi+\phi+i\chi)/\sqrt{2}$ (we use the linear representation throughout), a NGB (the CP-odd component) 
or the massless Majoron $\chi$ results. The Majoron mass can be generated once an explicit LNS breaking term, 
\beeq
{-\mathcal{L}_{\rm{LNB}} = -\frac{m^2}{4}(\Phi^2+{\Phi^*}^2)},
\label{soft-breaking}
\eeq
is introduced. 
Such a term breaks the $U(1)_L$ to $Z_2$ (under which $\Phi \rightarrow -\Phi$) and induces a mass for the Majoron as $m_\chi^2= m^2$. 
This mass term is however expected to be small (soft breaking) in t'Hooft's sense as $m \rightarrow 0$ limit enhances the overall symmetry of the framework. It has been argued \cite{Holman:1992us,Kallosh:1995hi,Kamionkowski:1992mf,Ghigna:1992iv,Rothstein:1992rh} that smallness of the pNGB mass can be related to explicit global symmetry (here LNS) breaking at large scale, such as at Planck scale ($M_{Pl}$) by gravity effects \cite{Giddings:1988cx,Coleman:1988tj,Rey:1989mg,Abbott:1989jw,Barbieri:1979hc,Akhmedov:1992hh}. To realize it, Planck suppressed non-renormalizable operators of the form $\Phi^{n}/M_{Pl}^{n-4} + h.c.$ with $n>4$ can be incorporated which are only soft (being non-renormalizable, such terms vanish in the limit $M_{Pl} \rightarrow \infty$) explicit symmetry breaking term(s). The mass of the Goldstone then follows after the spontaneous breaking of the global symmetry. 

Furthermore, the portal coupling can also be considered to be small ($\lambda_{H \Phi} \ll 1$) which is technically natural from the point of view of enhanced Poincare symmetry in the limit $\lambda_{H \Phi} \rightarrow 0$ \cite{Foot:2013hna,Coy:2022unt}. Considering a large hierarchy between the scales of lepton number and EW symmetry breaking along with tiny Higgs portal coupling, the mixing generated between the CP-even parts of $\Phi$ and $H$ turns out to be negligible which results into the following masses of the scalar fields as 
\beeq
{m_h^2 \simeq 2\lambda_H v^2,~ m_{\phi}^2 \simeq \lambda_\Phi v_\Phi^2,~ m_{\chi}^2 \simeq m^2}, 
\eeq
where $m_h$ is the mass of the SM Higgs boson as 125 GeV \cite{CMS:2012qbp}. 

\subsection*{Seesaw Mechanism and Interactions of Majorons}

Using the linear representation of the $U(1)_L$ breaking field $\Phi=\left( {v_\phi+\phi+i\chi}\right) /{\sqrt{2}}$, the following interaction of Majoron results via Eq. \ref{lag}, 
\beeq
-\mathcal{L} \supset \frac{if_i}{2\sqrt{2}}\chi \overline{\mathcal{N}^c_{R_i}} \mathcal{N}_{R_i} + h.c..
\label{Majoron-1}
\eeq
Furthermore, after the EWSB, a mixing between left and right-handed neutrinos takes place via the Dirac mass $m_D = y^\nu v /{\sqrt{2}}$ and 
consequently the mass terms involving $\nu_L$ and $N_R$ can be written as, 
\beeq -\mathcal{L}_{m}= \begin{pmatrix}
	\overline{\nu_L} & \overline{\mathcal{N}^c_R}
\end{pmatrix}\begin{pmatrix}
	0 & m_D \\
	m_D^T & M_R
\end{pmatrix} \begin{pmatrix}
\nu_L^c \\
\mathcal{N}_R
\end{pmatrix},  
\label{mass1}
\eeq
where the generation indices are suppressed. This yields the light and heavy neutrino mass matrices $m_{\nu}$ and $M$ ($\simeq M_R$, already diagonal) respectively in the rotated basis $(\tilde{\nu}_L^c~ ~\mathcal{N})^T$, where\footnote{$M_R$ being diagonal, $N$ coincides with $\mathcal{N}_R$.} $\tilde{\nu}_L^c = -i(\nu_L^c - \theta \mathcal{N}_R)$ with 
the active-sterile mixing matrix {$\theta = m_D^* M_R^{-1}$} and $ \mathcal{N} \equiv \mathcal{N}_R$. A further diagonalization of $m_{\nu}$ by the $U_{PMNS}$ \cite{ParticleDataGroup:2020ssz} leads to the diagonal mass matrix diag$(m_{\nu_j}, M_i)$ in the mass eigenstate basis $(n_j ~~\mathcal{N}_i)^T$. Defining the Majorana mass eigenstates of light ($\nu_j = n_j + n^c_j$) and heavy neutrinos ($N_i =\mathcal{N}_i+\mathcal{N}^c_i$) by $\nu_j$ and $N_i$ respectively, the interaction terms of Majoron (followed from Eq. \ref{Majoron-1}) with light and heavy neutrino mass eigenstates can be written as   
\bea
-\mathcal{L}_{\chi \nu \nu} &=&-\sum_{j,k}\mathcal{L}_{\chi \nu_j \nu_k} \nonumber\\
&=&-\frac{\chi }{2\sqrt{2}} \sum_{j,k} \left(\sum_{i}if_i \overline{\nu_j} P_R \nu_k V^T_{ji} V_{ik} +h.c. \right),\nonumber\\
\label{L3}
\eea
and
\bea
-\mathcal{L}_{\chi N \nu} &=&-\sum_{i,j}\mathcal{L}_{\chi N_i \nu_j} \nonumber \\
& = & \frac{\chi}{2\sqrt{2}}   \sum_{i,j} f_i \left(\overline{\nu_j} P_R N_i V^T_{ji}  + \overline{N_i}P_R \nu_j V_{ij}\right) + h.c.,\nonumber\\
\label{L4}
\eea
where $V = \theta^\dagger U$.

\subsection*{Production and Decay of Majorons}

As we are looking for Majoron as FIMP, the interactions of Majorons mentioned above are suggestive of its primary production via $N_i\to\chi \nu$, the associated decay width of which is given by,
\beeq
\Gamma_{N_i\to\chi \nu}= \frac{M_i^3}{32 \pi v_\phi^2}  \sum_{j=1,2,3} |V_{ij}|^2.
\label{d1}
\eeq 
Such a decay width can be shown to be approximately proportional to light neutrino mass ({$\Gamma_{N_i\to\chi \nu}\simeq \frac{1}{32\pi}\left(\frac{M_i}{v_\phi} \right)^2 \sum_j m_{\nu_j}  $}) \cite{Escudero:2016tzx} 
and hence suppressed. Also, this particular decay channel of $N_i$ opens up only after the electroweak symmetry breaking (EWSB) as the interaction in Eq. \ref{L4} contributing to this decay is proportional to the active-sterile neutrino mixing. Considering the RHNs are in thermal equilibrium at a temperature 
$T > M_i$ and the branching of this decay remains sizeable enough compared to other decays of $N_i$ ($e.g. ~N_i$ to SM ones via neutrino Yukawa interactions), the relic contribution of Majoron can be expressed as \cite{Hall:2009bx},
\beeq
\Omega_\chi h^2 \approx \frac{1.09\times 10^{27}}{g_\star^\mathcal{S} \sqrt{g_\star^\rho}}m_\chi \sum_{i}\frac{g_i\Gamma_{N_i\to\chi \nu}}{M_{i}^2},
\label{omega}
\eeq
where $g_\star^{\mathcal{S},\rho}$ is the effective number of degrees of freedom in the bath and $g_i=2$ denotes the internal spin degrees of freedom of RHN.

On the other hand, following Eq. \ref{L3}, Majoron can decay into light neutrinos having the decay width 
\beeq
\begin{multlined}
\Gamma_{\chi\to \nu \nu} = \frac{m_\chi}{16\pi v_\phi^2} \sum_{j} m_{\nu_j}^2\\
\simeq \frac{1}{10^{19}s}\left( \frac{m_\chi}{1~\textrm{MeV}}\right) \left( \frac{8\times10^8~\textrm{GeV}}{v_\phi}\right)^2 \left(\frac{\sum_{j}m_{\nu_j}^2}{2.6\times10^{-3}~\textrm{eV}^2} \right).
\end{multlined}
\label{d0}
\eeq
 In getting the sum of the light neutrino mass-squared in Eq. \ref{d0}, the best fit values of atmospheric and solar mass-splittings are used from neutrino oscillation data \cite{deSalas:2020pgw} where we consider a normal hierarchy of light neutrinos along with $m_{\nu_1}=0$. The stability criteria of Majoron to be a viable dark matter candidate as given by
\beeq
\Gamma^{-1}_{\chi\to \nu\nu} > \tau_{U}, 
\label{stability}
\eeq
needs to be satisfied, where $\tau_U$ is the lifetime of the universe  $\sim \mathcal{O}(10^{19})~\rm{sec}$ \cite{Arguelles:2022nbl}. 
Using Eqs. \ref{omega} and \ref{d0}, the above relation can therefore be employed to provide a limit on the relic contribution of Majoron as given by
 \beeq
\Omega_\chi h^2< 1.28\times10^{-9}\left(\frac{10^{19}~\rm{sec}}{\tau_U} \right). 
\eeq
The contribution turns out to be insignificant in making up the dark matter relic $\Omega_\chi h^2 \simeq 0.120\pm 0.001$ \cite{Planck:2018vyg}.  

Apart from the interactions with neutrinos mentioned in Eqs. \ref{L3} and \ref{L4}, Majorons also have interaction with the CP-even scalar 
$\phi$. However, in view of negligible portal coupling $\lambda_{H\Phi}$, this $\phi$ field can not be present in the thermal bath of the early Universe consisting of the SM fields. Since $\phi$ has no such important role to play, it is generally assumed that the $\phi$ field is heavy enough ($m_{\phi} \sim v_{\Phi}$, $i.e.$ with $\lambda_{\Phi} \sim \mathcal{O}(1)$) compared to the rest of the masses involved in the set-up and hence decoupled.  
Other possibility of $\chi$ production is from the process $H^{\dagger} {H} \rightarrow \chi \chi$ which is again proportional to tiny Higgs portal coupling and suppressed by $m_{\phi}$. This turns out to be a possibility of having heavy Majorons $\sim$ TeV \cite{Abe:2020dut}. Here, however, we plan to explore lighter Majorons as DM.

\section{The Proposal}

As we have discussed above, the minimal Majoron set-up cannot accommodate the required relic density of FIMP like DM as Majoron. In order 
to make it a viable option, we extend this set-up by introducing an explicit dimension-5 $U(1)_L$ breaking term  
\beeq
{-\mathcal{L}_{{\rm{d}}_5} =  
\frac{\alpha_i}{2\Lambda} \left[\Phi^2+(\Phi^*)^2\right]\overline{\mathcal{N}^c_{R_i}}{\mathcal{N}_{R_i}}+h.c.}, 
\label{d5}
\eeq
where $\Lambda$ is a cut-off scale. Here we follow a guideline considering that 
the explicit breaking of the global symmetry (LNS) takes place at some high scale ($\Lambda \leq M_{Pl}$) \cite{Draper:2022pvk,Cordova:2022rer} the manifestation of which is through the appearance of $non$-$renormalizable$ operators (of dimension 5 or more, suppressed by powers of $\Lambda$) only. Hence, the operator 
in Eq. \ref{d5} is the only leading order LNV operator\footnote{The usual dimension-5 operator contributing to light neutrino mass, $\frac{c_{ij}}{\Lambda} L_iHL_jH$, may also be present. However, assuming $c_{ij}$ to be small enough, we ignore this term for the analysis without loss of any generality.} involving RHNs and $\Phi$ field. For simplicity, we consider the coefficients $\alpha_i$ appearing in Eq. \ref{d5} to be $\mathcal{O}(1)$ and omit them for further discussion.

Note that the term of Eq. \ref{d5} being a higher order one, the $U(1)_L$ symmetry is only softly broken. This is similar to the origin of the Majoron mass of Eq. \ref{soft-breaking} where the smallness of $m$ is connected to higher order explicit LNV operator. As per our guideline above, such a mass term $m^2 (\Phi^2 + {\Phi^*}^2)$ may originate from a non-renormalizable explicit symmetry breaking term, $e.g. ~k \frac{|\Phi|^4}{\Lambda^2} (\Phi^2 + {\Phi^*}^2)$, which results $m^2 = k \langle \Phi \rangle^2 \left[ \frac{\langle \Phi \rangle}{\Lambda}\right]^2$ after spontaneous symmetry breaking and explains naturally the smallness of it as $ \langle \Phi \rangle \ll \Lambda$.

Such a term\footnote{Disappearance of any cross generation term is ensured in the mass-diagonal basis of RHNs.} also contributes to the mass of the RHNs, after the spontaneous breaking of $U(1)_L$ symmetry, as given by 
\beeq
M_{i} = v_\phi\left( \frac{f_i}{\sqrt{2}}+\frac{  v_\phi}{\Lambda}\right).
\label{RHN-mass}
\eeq 
Note that the inclusion of such a term carries the potential of generating a sizeable population of Majorons via the four-point interaction between Majorons and RHNs 
\beeq
-\mathcal{L}_{int} = \frac{1}{2\Lambda} \chi^2 \overline{N^c_i}N_i,
\label{effective-int}
\eeq 
induced by Eq. \ref{d5} after the $U(1)_L$ symmetry is spontaneously broken. This term contributes to the production of Majorons via $NN \rightarrow \chi \chi$ annihilation, somewhat similar to the UV freeze-in scenario as we elaborate below. Although suppressed by the cut-off scale $\Lambda$, this process turns out to be crucial in producing Majorons as it could remain in operation at a very early Universe (after the $U(1)_L$ symmetry breaking), contrary to the usual Majoron production from RHN decay $N \rightarrow \chi \nu$ being effective after the EWSB at temperature $T_{EW}$. 
\\
 
\subsection*{Majoron Production via annihilation}

As proposed above, the introduction of the effective interaction between RHNs and Majorons as in Eq. \ref{effective-int} opens up the new production channel for Majorons via annihilation of RHNs (see Fig. \ref{fynmann}) once the $U(1)_L$ symmetry is broken at a temperature\footnote{We consider the reheating temperature after inflation ($T_R$) to be bigger than $T_L$, though smaller than $\Lambda$.} $T_L \sim v_{\phi}$. It is interesting to note that the RHNs receive their masses from the spontaneous breaking of the $U(1)_L$ symmetry around this temperature only. Now, these RHNs may or may not be in thermal equilibrium. First, we consider the RHNs to be in equilibrium (as case[A]) and thereafter we investigate the situation when the abundance of RHNs at $T_L$ to be negligible (as case [B]) to begin with which is expected to be increased thereafter gradually by the neutrino Yukawa interaction. Following the discussion before, we assume the Higgs portal coupling $\lambda_{H\Phi}$ negligible and consider $\phi$ being heavy to be decoupled field. \\

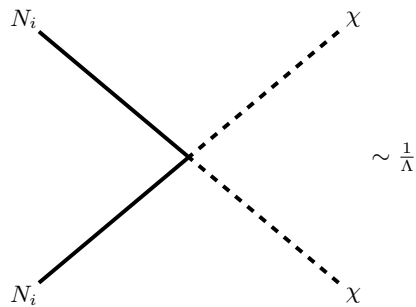
\begin{figure}[h!]	
	\begin{tikzpicture}[line width=1.5 pt, scale=1.3]
	\draw[fermionnoarrow] (-140:2)--(0,0);
	\draw[fermionnoarrow] (140:2)--(0,0);
	\node at (-140:2.2) {$N_i$};
	\node at (140:2.2) {$N_i$};
	\begin{scope}[shift={(0,0)}]
	\draw[scalarnoarrow] (-40:2)--(0,0);
	\draw[scalarnoarrow] (40:2)--(0,0);
	\node at (-40:2.2) {$\chi$};
	\node at (40:2.2) {$\chi$};
	\node at (2.1,0) {$\sim\frac{1}{\Lambda} $};
	\end{scope}
	\end{tikzpicture} 
	\caption{Annihilations of RHNs to Majorons}
	\label{fynmann}
\end{figure}

To study the evolution of the yield of Majoron as DM, we use the Boltzmann equation written in terms of the temperature $(T)$ and the comoving number density of $\chi$ as $Y_\chi=n_\chi/\mathcal{S}$ (here $\mathcal{S}$ is the entropy density). A general form of this equation (irrespective of whether $N_i$ and/or $\chi$ are in equilibrium) can be written as follows \cite{Kolb:1990vq,Chianese:2019epo}\\
\beeq
\begin{multlined}
	\mathcal{H}T\frac{dY_\chi}{dT}= -2{\mathcal{S}}\sum_{i}\big\langle\sigma v\big\rangle_{N_i N_i\to \chi\chi} Y_{N_i}^{eq^2}\left(\frac{Y_{N_i}^2}{Y_{N_i}^{eq^2}}-\frac{Y_\chi^2}{Y_\chi^{eq^2}} \right)\\
	-\sum_{i,j}\big\langle\Gamma_{N_i\to \chi \nu_j}\big\rangle Y_{N_i}^{eq}\left(\frac{Y_{N_i}}{Y_{N_i}^{eq}}-\frac{Y_\chi Y_{\nu_j}}{Y_\chi^{eq} Y_{\nu_j}^{eq}} \right)~\theta(T_{EW}-T). 
	\label{BEbasic}
\end{multlined}
\eeq
Here, $\mathcal{H}$ is the Hubble rate given by $\mathcal{H} =1.66\sqrt{g^\rho_\star(T)} T^2/M_{Pl}$ and $\mathcal{S}=\frac{2\pi^2}{45}g^{\mathcal{S}}_\star(T)$, where $g^{\mathcal{S}}_\star(T)$ and $g^\rho_\star(T)$ are the effective degrees of freedoms of relativistic species at temperature $T$ having $M_{Pl}=1.22\times 10^{19}$ GeV as the Planck mass. At high temperature, $g^{\mathcal{S}}_\star(T)=g^\rho_\star(T)=106.75$ follows from  SM particle content. The yield of a particle species $a$ in thermal equilibrium ($Y_{a}^{eq}$) is given by,
 \beeq
 Y_{a}^{eq} (T)= \frac{45 g_a}{4\pi^4 g^\mathcal{S}_{\star}(T)}\left( \frac{m_a}{T}\right)^2 K_2\left(\frac{m_a}{T} \right), 
 \eeq 
 where $g_a=1(2)$ (for scalar (fermion)) and $m_a$ are the internal degrees of freedom and mass of the particle species respectively. $K_2$ is the modified Bessel's function of the second kind.  $\big\langle\sigma v\big\rangle$ and $\big\langle\Gamma\big\rangle$ are thermally averaged cross section and decay width respectively, the estimate of which are described below in the following section. The relic density of $\chi$ is obtained after substituting the freeze-in abundance $Y_\chi(T_0)$ in,
\beeq
 \Omega_\chi h^2=2.755\times10^8\left(\frac{m_\chi}{\rm{GeV}} \right)Y_\chi(T_0), 
 \eeq
 where $T_0$ is the present temperature. 
 
 \section{Dark Matter phenomenology}
 
 We now study the role of the higher dimensional interaction of Eq. \ref{d5} in obtaining the relic density of the Majoron field $\chi$ which is expected to play the role of freeze-in DM. From the discussion of the previous section, we understand that this operator contributes to the production of Majorons from the annihilations of RHNs. The matrix element squared for the annihilation process $N_{i}N_{i}\to \chi\chi$ is given by,  
\beeq
|\overline{M}|^2_{N_{i}N_{i}\to \chi\chi}= \frac{1 }{\Lambda^2}\left(s-4M_i^2 \right),
\label{ampl}
\eeq
leading to the cross section
\beeq
\sigma_{N_{i}N_{i}\to \chi\chi}=\frac{1}{16\pi s}\sqrt{\frac{s-4m_{\chi}^2}{s-4M_{{i}}^2}} |\overline{M}|^2_{N_{i}N_{i}\to \chi\chi},
\label{cross-sec}
\eeq
where $\sqrt{s}$ is the centre of mass energy of the process at a temperature $T$. 

We consider $U(1)_L$ symmetry to be broken below the reheating temperature, $T_R$ indicating $T_L < T_R$. Since the origin of the Majoron field is intertwined with the breaking of this global symmetry, the initial temperature for studying the $\chi$ abundance through the Boltzmann equation(s) (via Eq. \ref{BEbasic}) can be considered as $T_L \sim v_{\phi}$ \cite{Berezinsky:1993fm} instead of the reheating temperature as in UV freeze-in scenario \cite{Elahi:2014fsa}. In this way, we do not need to invoke a parameter outside the framework (such as $T_R$) in the parameter space scan. However, without any prior knowledge on $T_R$ and keeping in mind that RHNs get their masses at $T_L$, RHNs may or may not be in equilibrium at this temperature. 

With the above points in mind, below we proceed for the evaluation of the freeze-in relic density of Majoron as DM corresponding to two cases: [A] RHNs are in thermal equilibrium and [B] RHNs are not in thermal equilibrium at $T_L$ and find out the relevant parameter space. The initial abundance of the $\chi$ particle is considered to be zero at this temperature $T_L$ in both cases. As we have already seen in section \ref{section2} that the contribution $N_i\to\chi\nu$ channel to the $\chi$ abundance is almost negligible, we 
proceed further without it.

\subsection{RHNs: in thermal bath at $T_L$}

With the consideration that the RHNs are already in thermal equilibrium and abundance of $\chi$ remains vanishing at $T_L$, the Majorons can be produced from the annihilation via dimension-5 operator whose yield can be estimated by the following Boltzmann equation for $\chi$
 \beeq
\begin{multlined}
 \frac{dY_\chi}{dT} \simeq  -\frac{2\mathcal{S}}{\mathcal{H}T} \sum_{i} \big\langle\sigma v\big\rangle_{N_iN_i\to\chi\chi} Y_{N_{i}}^{eq^2} \left[ 1-\frac{Y_\chi^2}{Y_{\chi}^{eq^2}}\right] ,
\end{multlined}
\label{BE1}
\eeq
\\
\noindent which is a simplified form of  Eq. \ref{BEbasic}. First term in $r.h.s$ of Eq. \ref{BE1} refers to the production 
of $\chi$ due to the annihilation of RHNs where the factor 2 in front arises due to production of two Majorons in the final state. The second term in the $r.h.s$ related to the back reaction $\chi\chi\to N_i N_i ~(\textrm{and}~\chi\nu\to N_i)$ is not important as $\chi$ starts with a negligible abundance in general. However, for light Majoron (of $\sim$ keV mass), the $\chi$ abundance near freeze-in temperature can be comparable (though remains smaller) to its equilibrium abundance where this term would be significant. On the other hand, the equilibrium abundance of RHNs is given by 
\beeq
 Y_{N_{i}}^{eq} (T)=\frac{45 g_{i}}{4\pi^4 g^\mathcal{S}_{\star}(T)}\left( \frac{M_i}{T}\right)^2 K_2\left(\frac{M_i}{T} \right), 
 \eeq 
 \\
\noindent with $g_i=2$. The thermally averaged cross-section appearing in Eq. {\ref{BE1}}, can be expressed as \cite{Gondolo:1990dk}
\beeq
\begin{multlined}
\big\langle\sigma v\big\rangle_{N_iN_i\to\chi\chi}=
	\frac{1}{8M_i^4TK_2^2(M_i/T)}\times\\
	\int_{4M_i^2}^{\infty}\sigma_{N_iN_i\to\chi\chi} (s-4M_i^2)\sqrt{s}K_1\left( { \sqrt{s}}/{T} \right)  ds. 
\end{multlined}
\label{sigmav}
\eeq
By integrating Eq. \ref{BE1} from $T_L \sim v_{\phi}$ (from where $\chi$ production takes place) to $T_0$ (today), we can obtain the yield of $\chi$.


Note that the parameters involved in analyzing the Majoron or $\chi$ production are: $f_1, f_2$ (related to RHN masses $M_1, M_2$), $v_\phi, m_{\chi}$ and the cut-off scale $\Lambda$. For simplicity, we consider the couplings (real) $f_1, f_2$ to be same ($f_1 = f_2 = f$) and fix $f$ at some natural value, $f = 0.01$ so that the degenerate RHN masses $M_i = M$ always remain below $v_\phi$. Thereafter, we employ the Boltzmann equation (Eq. \ref{BE1}) to perform a scan over the parameter space so as to satisfy the DM relic density by the $\chi$ field. In doing so, we implicitly consider $\Lambda$ being the cut-off scale is bigger than $v_{\phi}$, however, bounded by the largest scale $M_{Pl}$. Furthermore, the stability criteria via Eq. \ref{stability} must be satisfied. We also incorporate an upper limit on $v_{\phi}$ as the maximum value for the reheating temperature is generally considered to be $T^{\rm{max}}_R\sim10^{15}$ GeV \cite{Haque:2020zco} ($v_{\phi} < T_R$ as stated earlier). 

Our findings are displayed in Fig. \ref{PS1-th} by the coloured region corresponding to the correct relic satisfaction in the $m_{\chi} - v_{\phi}$ plane. The colours in this relic satisfied $m_{\chi} - v_{\phi}$ space are indicative of the corresponding $\Lambda$ value in the right-side colour bar. The gradient of the colour (from blue to dark red side) is increased with the increase in $\Lambda$ which saturates at $M_{Pl}$. The lower boundary of the allowed parameter space follows from the stability constraint while the top-left disallowed region (white patch) is limited by a specific choice $v_{\phi}/{\Lambda} < f/\sqrt{2}$, so that the dimension-5 operator's contribution to the RHN mass (see Eq. \ref{RHN-mass}) remains sub-dominant. On the other hand, the top-right disallowed region corresponds to $\Lambda > M_{Pl}$. 
\begin{figure}[!hbt]
	\centering
	\includegraphics[scale=0.5]{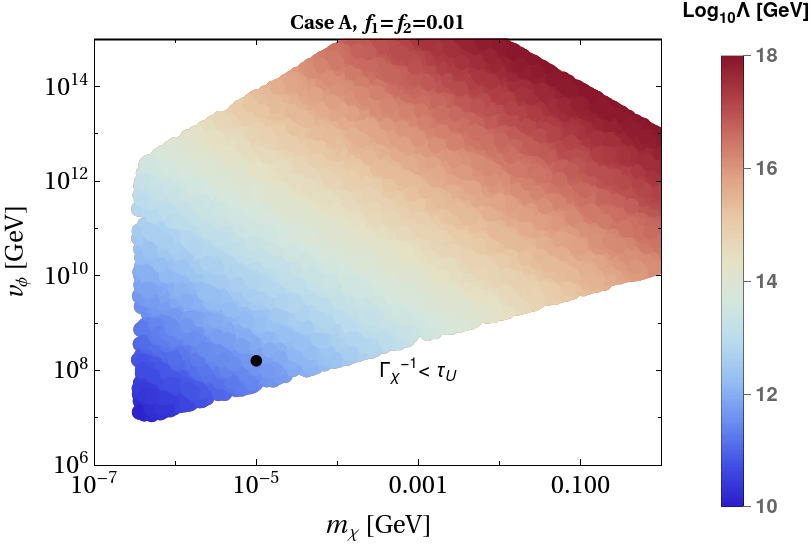}
	\caption{Case A (degenerate RHNs are in thermal equilibrium at $T_L$): relic satisfied parameter space represented by coloured region in $v_\phi$ vs $m_\chi$ plane, while $\Lambda$ are in colour side-bar. The black dot indicates the BP considered in Table \ref{BP-table}.}
	\label{PS1-th}
\end{figure}
The leftmost region of the allowed parameter space is bounded by the model-independent limit \cite{Elahi:2014fsa} on the Majoron mass $m_{\chi} < \mathcal{O}(0.1)$ keV, signifying that the Majorons can never be able to reach thermal equilibrium. We find for Majoron as light dark matter in the range 0.1 keV - 1 GeV, the Lepton number breaking scale $v_{\phi}$ and $\Lambda$ fall in the following range,
\bea
m_{\chi}:  0.4~ {\rm{keV}} - 1~ {\rm{GeV}};\nonumber \\
v_\phi: 10^6~ {\rm{GeV}} - 10^{15} ~{\rm{GeV}}; \nonumber \\
\Lambda: 10^{10} ~{\rm{GeV}} - 10^{18}~ {\rm{GeV}}. 
\nonumber
\eea

In order to demonstrate the freeze-in of the Majoron, we include Fig. \ref{BP1-th} where the abundance of the $\chi$ field is presented as a function of temperature $T$. This corresponds to a specific benchmark set of points (BP) from the parameter space as included in Table \ref{BP-table} (indicated by a dark dot on the allowed parameter space of Fig. \ref{PS1-th}).
Note that as the RHNs are in thermal equilibrium from the beginning (indicated by the blue line), the maximum production of $\chi$ (in the orange line) from the RHN annihilation takes place near the temperature $T_L$ itself, where the $U(1)_L$ is broken. Such a behaviour is similar to the usual UV freeze-in scenario. 
\begin{figure}[h!]
	\centering
	\includegraphics[scale=0.35]{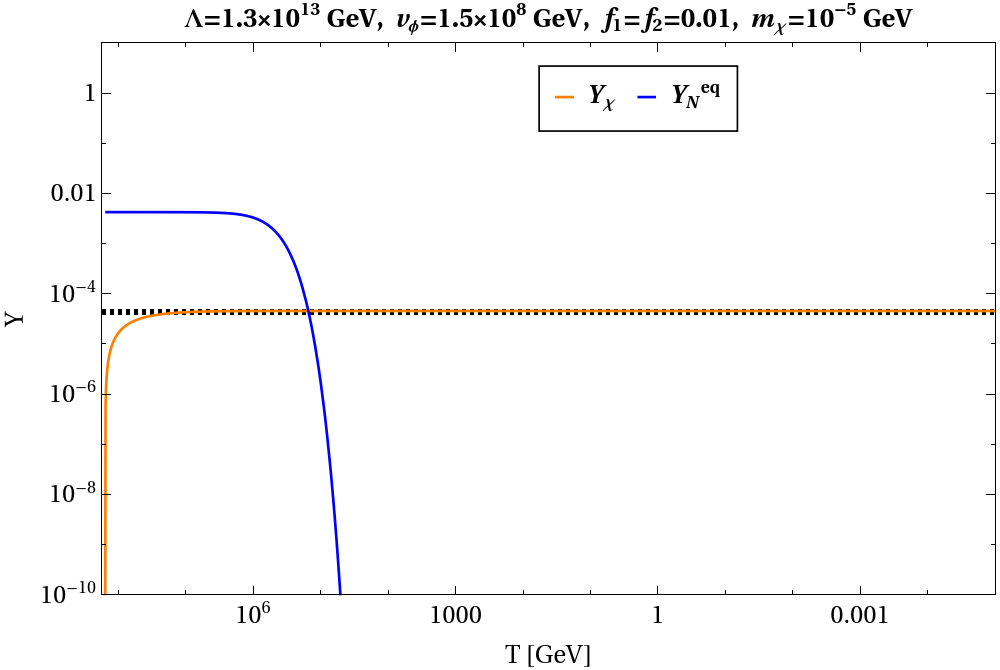}
	\caption{Evolution of $Y_\chi$ (solid orange), $Y_{N_i}=Y_{N_i}^{eq}$ (solid blue) against $T$ (GeV) are shown corresponding to the BP for case A from Table \ref{BP-table} with degenerate RHNs $(f_1=f_2=0.01)$. The grid line indicates the correct yield for the relic satisfaction.}
	\label{BP1-th}
\end{figure}
The horizontal grid line corresponds to the yield of $\chi$ producing the correct relic abundance for $m_{\chi} =10$ keV. Note that this result remains unchanged even if we consider non-degenerate RHNs. Since both the RHNs (even if $M_1 \neq M_2$) are considered to be in thermal equilibrium at $T_L$, it does not affect the production of $\chi$ from $N_i$ annihilation which happens to be around $T_L$. Next, we proceed to investigate the situation when RHNs are not in the thermal bath. 
\begin{center}
\begin{table}[h]
\begin{tabular}{ |c|c|c|c|c| } 
	\hline
	BP & $\Lambda$ (GeV) & $v_\phi$ (GeV) & $m_\chi$ (GeV) \\
	\hline
	\hline
	case A & $1.3\times10^{13}$ & $1.5\times 10^8$ & $10^{-5}$\\ 
	\hline
	case B.1 & $10^{12}$ & $1.6\times 10^8$ & $10^{-5}$\\
	\hline	
\end{tabular} 
\caption{Benchmark Point (B.P.) for case A and case B.1. For both cases $f_1=f_2=0.01$ is considered.}
\label{BP-table}
\end{table}
\end{center}

\subsection{RHNs: not part of the thermal bath at $T_L$}

In this sub-section, we consider the case where RHNs are not in thermal equilibrium\footnote{Recall that RHNs become massive at $T_L$ only.} at $T_L$. Hence, we do not expect an identical behaviour for the Majoron production in terms of its dominant production near $T_L$. Rather, first the RHNs are being produced from the thermal bath by the inverse decay process: $L H \rightarrow N_i$ and consequently the Majoron abundance should be increased from the annihilations of the RHNs gradually. 

Initially, we presume negligible abundance of both the $\chi$ and the $N_i$ at $T_L$. Then we solve the coupled Boltzmann equations for Majoron as well as RHNs 
$(N_i)$ as given by,
\bea 
 \frac{dY_{N_i}}{dT}&=& \frac{2\mathcal{S}}{\mathcal{H}T}\big\langle\sigma v\big\rangle_{N_iN_i\to\chi\chi} \left(Y_{N_i}^2-\frac{(Y_{N_i}^{eq})^2}{(Y_{\chi}^{eq})^2}Y_{\chi}^2 \right)\nonumber\\
 &&+\frac{1}{\mathcal{H}T}\big\langle\Gamma_{N_i\to LH}\big\rangle\left(Y_{N_{i}}- {Y_{N_i}^{eq}} \right),
 \eea
 \bea
 \frac{dY_\chi}{dT}&=& -\frac{2\mathcal{S}}{\mathcal{H}T}\sum_{i}\big\langle\sigma v\big\rangle_{N_iN_i\to\chi\chi} \left(Y_{N_i}^2-\frac{(Y_{N_i}^{eq})^2}{(Y_{\chi}^{eq})^2}Y_{\chi}^2 \right),\nonumber\\  
 \label{BE3}
 \eea 
where, $\big\langle\sigma v\big\rangle_{N_iN_i\to\chi\chi}$ follows from Eq. \ref{sigmav} and $\big\langle\Gamma_{N_i\to LH}\big\rangle$ is the thermal averaged decay width of $N_i\to LH$ expressed as,
\beeq
\big\langle\Gamma_{N_i\to LH}\big\rangle=\Gamma_{N_i\to LH }\frac{K_1(M_i/T)}{K_2(M_i/T)},
\eeq
with 
\beeq
\Gamma_{N_{i}\to LH}=\sum_{\alpha}\frac{|y^\nu_{\alpha i}|^2 M_{i}}{8\pi}.
\eeq 

Note that in this case, the neutrino Yukawa coupling $y^\nu_{\alpha i}$ plays important role in the production of RHNs from the thermal bath which in turn affects the production of Majorons. The structure of $y^\nu_{\alpha i}$ can be extracted using the Casas-Ibarra (CI) parametrization \cite{Casas:2001sr} as:
\beeq
y^\nu =i\frac{\sqrt{2}}{v}U^\dagger D_{\sqrt{m_\nu}} R D_{\sqrt{M_R}},
\label{CI}
\eeq
where $D_{\sqrt{m_\nu}}$ $(D_{\sqrt{M_R}})$ is the squared-root of the diagonal active neutrino (RHN) mass matrix, and $U$ is the PMNS mixing matrix consisting of the three mixing angles and a CP-violating phase. Here, $R$ is a complex orthogonal matrix. We take the lightest active neutrino mass to be zero so as to define the $D_{m_\nu}$ from the best fit values \cite{Esteban:2020cvm,Esteban:2016qun} of the two mass squared differences from neutrino oscillation data \cite{Esteban:2020cvm}. The fitted values \cite{Esteban:2020cvm,Esteban:2016qun} of mixing angles and CP phase can be used to define $U$. As we are working with two RHNs, the structure of the $R$ matrix is chosen here as \cite{Ibarra:2003up},
\beeq
R=\begin{pmatrix}
	0 & 0\\
	\textrm{cos}~\theta_R & 	\textrm{sin}~\theta_R\\
	-\textrm{sin}~\theta_R & 	\textrm{cos}~\theta_R
\end{pmatrix},
\eeq 
where $\theta_R$ is in general a complex angle. With these inputs along with $v = 246$ GeV and a specific choice of $\theta_R$, one can define the neutrino Yukawa coupling $y^\nu$ which is used in evaluating the decay width of RHNs appearing in the set of Boltzmann equations. 

\subsubsection{With Degenerate RHNs}

\begin{figure}[h!]
	\centering
	\includegraphics[scale=0.42]{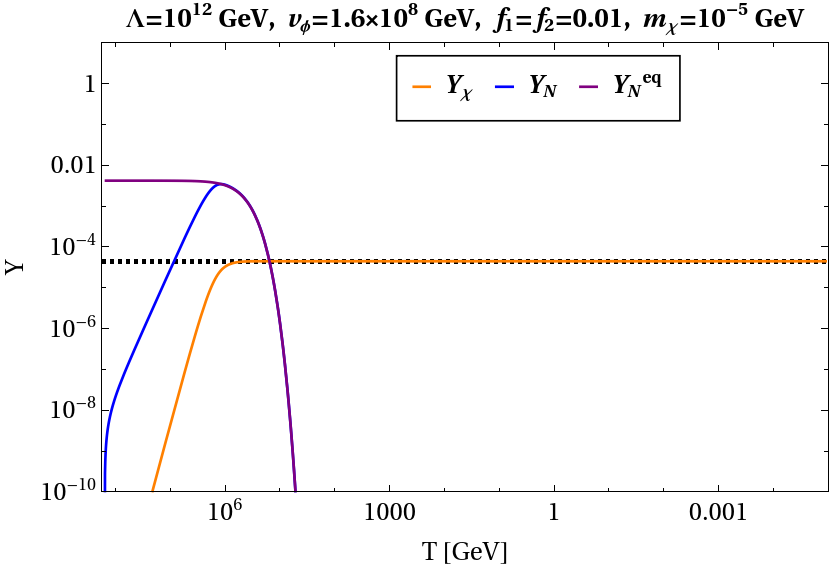}	
	\caption{Evolution of $Y_\chi$ (solid orange), $Y_{N_i}$ (solid blue) and $Y_{N_i}^{eq}$ (solid purple) against $T$ (GeV) are shown corresponding to the BP for case B.1 from Table \ref{BP-table} with degenerate RHNs and $\theta_R = \pi/4$. The black dotted line indicates the correct yield for the relic satisfaction.}
	\label{BP1-B}
	\end{figure} 	
	In order to check the expectations for the $\chi$ abundance and its freeze-in scenario compared to case A, we provide Fig. \ref{BP1-B} where benchmark set of parameters are assigned as per Table \ref{BP-table}, second row (namely, case B.1). Furthermore, in line with the discussion for case A, here also we consider the degenerate RHNs (by choosing same values for $f_1$ and $f_2$ as $f$).
We find that starting with negligible initial abundance, the yield of $N_i$ (indicated by the blue line) has gradually increased via the inverse decay and finally around $T \sim M$, it catches the equilibrium abundance (purple line), thanks to the Yukawa coupling $(y^\nu_{\alpha i})$. Note that there exists an additional parameter $\theta_R$ in this case as $y^{\nu}$ depends upon it. We fix it\footnote{This particular choice of $\theta_R$ with degenerate RHN mass corresponds to the same decay widths for both the RHNs, hence $N_1$ and $N_2$ reach equilibrium simultaneously. A general picture is demonstrated in the next subsection.} to be as $\pi/4$. Similar to case A, production of $\chi$ occurs from the four-point interactions between the RHNs and $\chi$ (after $U(1)_L$ symmetry breaking). However, contrary to case A, the production of $\chi$ here extends over a period from $T_L$ to $T \sim M$.  The freeze-in of $\chi$ occurs near the point when $N_i$ enters into thermal equilibrium. 

 
 
  \begin{figure}[!hbt]
 	\centering
 	\includegraphics[scale=0.5]{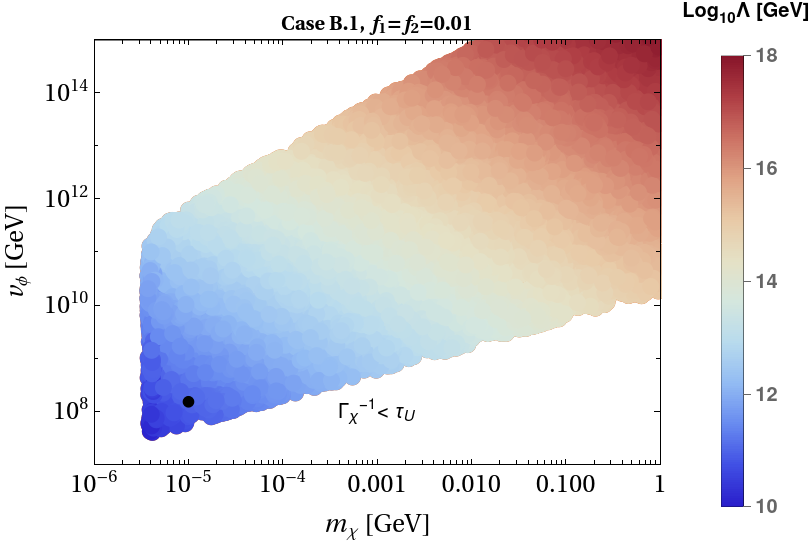}
 	\caption{Case B.1 (degenerate RHNs are not in thermal equilibrium at $T_L$): relic satisfied parameter space represented by coloured region in $v_\phi$ vs $m_\chi$ plane, while $\Lambda$ are in colour side-bar. The black dot indicates the BP considered in Table \ref{BP-table}.}
 	\label{PS1}
 \end{figure} 

 We then perform a parameter space scan with fixed $f = 0.01$ and $\theta_R = \pi/4$ which produced the correct relic density by Majoron field which is provided in Fig. \ref{PS1}. We find the following range of parameters 
\bea
m_{\chi}: 3~ {\rm{keV}} - 1~ {\rm{GeV}};\nonumber \\
v_\phi: 10^7~ {\rm{GeV}} - 10^{15} ~{\rm{GeV}}; \nonumber \\
\Lambda: 10^{10} ~{\rm{GeV}} - 10^{18}~ {\rm{GeV}}. 
\nonumber
\eea
allowed from the relic satisfaction point of view. Similar to case A, the top-left boundary appears due to the consideration $v_{\phi}/\Lambda < f/{\sqrt{2}}$ where we have restricted ourselves for $m_{\chi}$ below 1 GeV. The other imposed condition $v_{\phi} < T_R^{\rm{max}} \sim 10^{15}$ GeV represents the upper limit of $v_\phi$ in the parameter space as reflected in Fig. \ref{PS1}. Contrary to case A, here the leftmost boundary of the parameter space has shifted to $m_{\chi} \sim 3$ keV. This shift arises in case B.1 as a result of delayed thermal equilibration of RHNs which marks the efficient production of 
$\chi$ (from RHN annihilations) near $T \sim M_i$ compared to case A. Moreover, for the same reason we notice that corresponding to a similar set of parameters ($m_{\chi}, v_{\phi}$) in case A, we require here a relatively smaller $\Lambda$ to satisfy the correct relic (refer to Table \ref{BP-table}). This is reflected in Fig. \ref{PS1} where $\Lambda$ remains well within its upper limit $M_{Pl}$ so as not to exhibit any disallowed region 
(which exists for Fig. \ref{PS1-th} beyond $m_{\chi} >$ 2 MeV). We also find that there exists a lower limit on the spontaneous symmetry breaking scale, $v_{\phi} > 4 \times 10^7$ GeV (bottom leftmost corner in the Fig. \ref{PS1}) \cite{Alvi:2022aam}.

\subsubsection{With Non-Degenerate RHNs}
\begin{figure}[]
	\centering
	\includegraphics[scale=0.55]{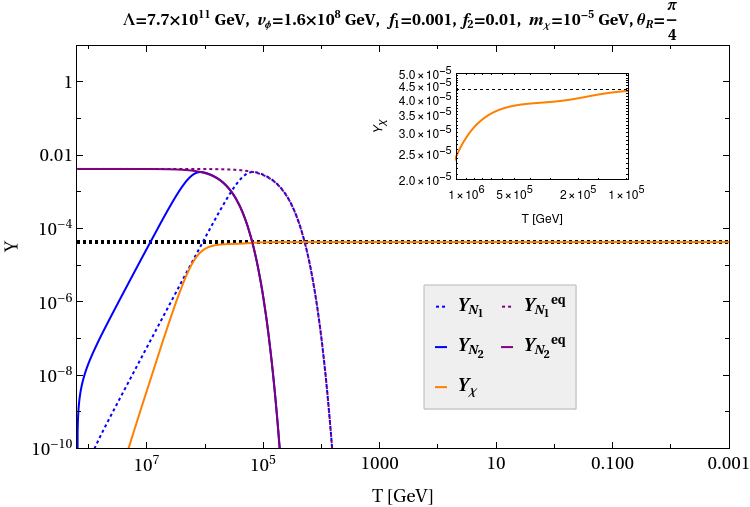}	
	\caption{Evolution of $Y_\chi$ (solid orange), $Y_{N_1}$ (dashed blue) and $Y_{N_2}$ (solid blue) along with $Y_{N_1}^{eq}$ (dashed purple) and $Y_{N_2}^{eq}$ (solid purple) against $T$ (GeV) are shown corresponding to the BP from 1st row, Table \ref{BP-table2} (with $f_1=0.001,~f_2=0.01$). The black dotted line indicates the correct yield for the relic satisfaction. In the inset, $Y_\chi$ is shown between $T\sim M_1 $ and $T\sim M_2$.}
	\label{BP1-B2}
\end{figure}
So far, we consider the case with degenerate RHNs (taking $f_1=f_2$). However, in a more general case where the two RHNs are not of the same mass ($i.e. ~f_1\neq f_2$), it will be interesting to investigate the behaviour of DM yield especially in the context of case B, where RHNs are not in thermal equilibrium at $T\sim T_L$. To elaborate it further, we start with 
same choices for $v_\phi, m_\chi, \theta_R, f_2$ related to case B.1 as in Table \ref{BP-table}, whereas differ in choosing $f_1=0.001$ $(i.e. ~M_1<M_2)$. It turns out that this non-degeneracy leads to a relatively lower $\Lambda=7.7\times 10^{11}$ GeV (in comparison to $\Lambda=10^{12}~\rm{GeV}$ with $f_1 = f_2 =0.01$, see Fig. \ref{BP1-B}) in order to satisfy the correct relic for $\chi$. The associated evolution of the Majoron abundance is shown in Fig. \ref{BP1-B2}. 
This new set of parameters is indicated in first row of Table \ref{BP-table2}. As observed in Fig. \ref{BP1-B2}, the production of $\chi$ initially takes place from $N_2$ annihilation during the epoch when  $N_2$ (with heavier mass) gradually approaches its equilibrium density (at $T \sim M_2$) due to inverse decay process. However, the production of Majoron continues beyond this temperature as the production from $N_1$ annihilation sets in. This introduces a kink in the yield of Majoron which is demonstrated in the inset of the figure. Finally the yield of $\chi$ freezes in when the lighter RHN ($N_1$) enters equilibrium at $T\sim M_1$.
\begin{widetext}
	\begin{center}
		\begin{table}[h]
			\begin{tabular}{ |c|c|c|c|c|c|c| } 
				\hline
				 $v_\phi$ (GeV)& $m_\chi$ (GeV) & $f_1$ & $f_2$  & $\theta_R$ &$\Lambda$ (GeV)   \\
				\hline
				\hline
			 $ 1.6\times10^8$& $10^{-5}$ & $0.001$ & $0.01$  & $\frac{\pi}{4}$ & 	$7.7\times10^{11}$ \\ 
				\hline
				 $1.5\times 10^{13}$ &$2.5\times10^{-3}$ & $0.009$ & $0.1$ &  $2.5+1.463i $ &$2\times10^{16}$ \\
				\hline	
			\end{tabular} 
			\caption{Benchmark Points (BP) for case B.2 with non-degenerate RHNs.}
			\label{BP-table2}
		\end{table}
	\end{center} 
\end{widetext}


 We incorporate a set of benchmark parameters in the second row of Table \ref{BP-table2} to mark the dependence on the other parameter $\theta_R$ (which in general can be complex) through $y^{\nu}$ matrix. As in this case, initially, the abundance of RHNs is negligible and it enters equilibrium via the neutrino Yukawa interaction, $\theta_R$ plays important role in determining the strength of this interaction and in a way affects the production of Majoron as well from RHN annihilation. Here, we consider a complex\footnote{Note that a complex $\theta_R$ can be the source of CP violation in leptogenesis. This particular choice of $\theta_R$ is related to correct baryon asymmetry generation via leptogenesis with $M_{1(2)}$ considered here \cite{Bhattacharya:2021jli,Datta:2021elq}.} $\theta_R = 2.5 + 1.463i$. The corresponding evolution of $N_{1(2)}$ and Majoron abundances are shown in Fig. \ref{BP1-B3} by blue dotted (solid) and orange lines respectively. To exhibit the effect of 
$\theta_R$ on the abundance of $\chi$, we now choose a real $\theta_R = 2.5$ while keeping all other parameters unchanged and exhibit the evolutions of RHNs and $\chi$ (by red dotted (solid) and green lines respectively) in the same figure. As we can see, this makes the Majoron under-abundant.

\begin{figure}[t!]
	\centering
	\includegraphics[scale=0.55]{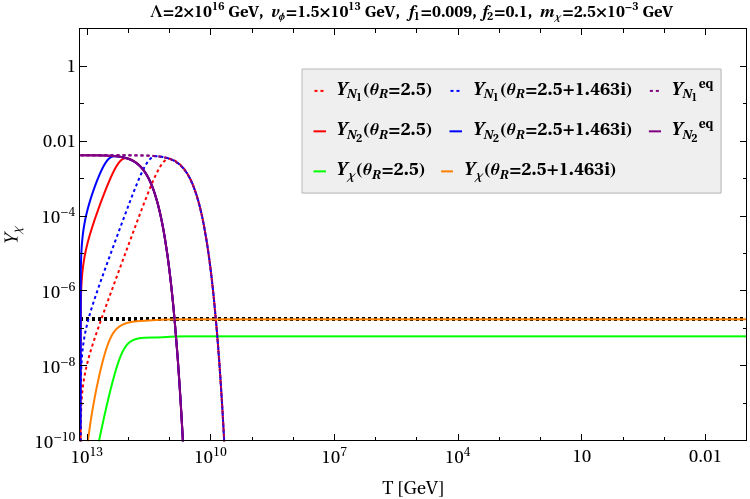}	
	\caption{ Evolution of $Y_\chi$, $Y_{N_1}$ and $Y_{N_2}$ for two different choices of $\theta_R~(2.5~\textrm{and}~2.5+1.463i)$ along with $Y_{N_1}^{eq}$ and $Y_{N_2}^{eq}$ are plotted against $T$ (GeV). All other parameters correspond to the BP from Table \ref{BP-table2}, second row. The black dotted line indicates the correct yield for the relic satisfaction.}
	\label{BP1-B3}
\end{figure}

\begin{figure}[t!]
	\centering
	\includegraphics[scale=0.5]{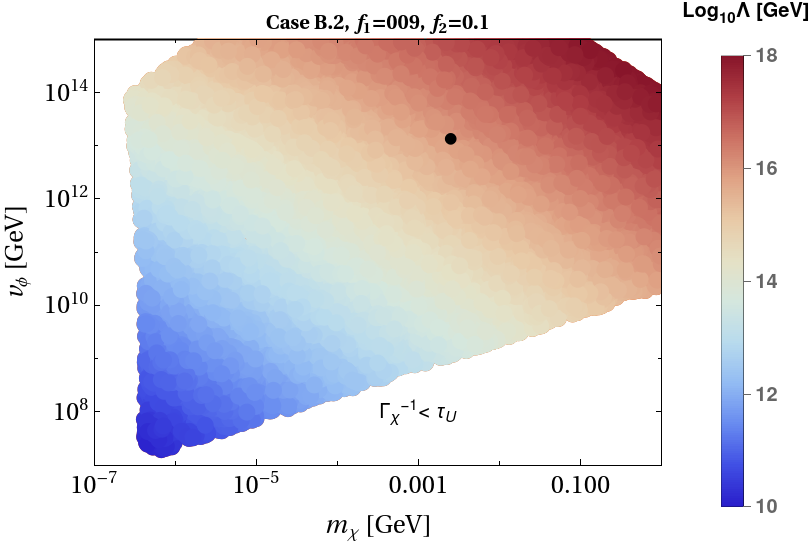}
	\caption{Case B.2 (non-degenerate RHNs are not in thermal equilibrium at $T_L$): relic satisfied parameter space represented by coloured region in $v_\phi$ vs $m_\chi$ plane, while $\Lambda$ are in colour side-bar. Here, $\theta_R$ is kept fixed as  $2.5+1.463i$. The black dot indicates the BP considered in Table \ref{BP-table2}, second row.}
	\label{PS2}
\end{figure}

With this particular choice of $\theta_R~(= 2.5+1.463i)$ and $f_1 = 0.009, f_2 = 0.1$, we represent the result of  another parameter space scan in Fig. \ref{PS2} in $v_{\phi} - m_{\chi}$ plane. We note that the allowed parameter space 
becomes broadened in comparison with Fig. \ref{PS1-th} and \ref{PS1}. This is due to the inclusion of additional parameter in the set-up in the form of different $f_1$ and $f_2$ along with complex $\theta_R$. The role of a complex angle $\theta_R$ can be significant in the context of the model-dependent constraints on Majoron (such as constraint on $\chi\to\gamma\gamma$ channel etc.) as we discuss it in the following section.
 
 \section{Constraints}
 
 As already discussed, the presence of the decay channel $\chi\to\nu \nu$ can make the Majoron unstable. Hence, in order to make the Majoron a viable dark matter, the condition $\tau_\chi> \tau_{U}$ needs to be employed which translates into the following expression from the recent analysis \cite{Alvi:2022aam} using \textit{Planck} \cite{Planck:2019nip} with CMB lensing \cite{Planck:2018lbu} and BAO data as \cite{Alvi:2022aam},
{\be
\Gamma(\chi\to\nu\nu)< 1.29\times 10^{-19}~ \textrm{s}^{-1}. 
\nonumber
\ee}
We have already incorporated this limit on the parameter space. We now turn our attention to other constraints applicable to the parameter space obtained in our set-up originating from supernova, cosmology as well as some indirect search experiments. 

The same tree-level coupling of Majoron with light neutrino mass eigenstates (see Eq. \ref{L3}) may also experience a constraint from supernova (SN). The basic idea stems from the fact that inside the supernova core, Majorons can be produced through the process $\nu \nu \to \chi$ (mainly $\nu_e$ takes part inside the SN core). Provided the neutrino-Majoron coupling parametrized by $g_{\chi \nu \nu} \equiv \sum_i m_{\nu_i}/{2v_{\phi}}$ is large, such Majorons would affect \cite{Brune:2018sab,Escudero:2020ped,Heurtier:2016otg,Farzan:2002wx,Akita:2022etk,Fiorillo:2022cdq} the neutrino signal emitted from a core collapsing SN which is otherwise consistent with the binding energy released during SN explosion as measured for SN1987A \cite{Kamiokande-II:1987idp,Bionta:1987qt}. Such a consideration leads to a disallowed range of the coupling $g_{\chi \nu \nu}$: ($10^{-7} - 10^{-5}$) for $m_{\chi} <$ 10 MeV and ($10^{-9} - 10^{-6}$) for $m_{\chi} <$ 200 MeV, as evaluated in \cite{Escudero:2020ped}. However, our parameter space scan indicates that $v_{\phi} \gtrsim \mathcal{O}(10^{7})$ GeV and accordingly  $g_{\chi \nu \nu} \lesssim 2.5\times 10^{-18}$ (where $\sum_i m_{\nu_i} \simeq 0.05$ eV is inserted). Hence, it remains way below the limit from supernova constraint. 

Also, this Majoron decay into light neutrinos may produce observable monochromatic neutrino fluxes which can be analyzed in dedicated experiments such as Borexino (testable at $4$ MeV $<m_\chi<60~$MeV) \cite{Borexino:2010zht}, KamLAND (testable at $16$ MeV $<m_\chi<60~$MeV) \cite{KamLAND:2011bnd}, SK (within $30$ MeV $<m_\chi<200~$MeV \cite{Super-Kamiokande:2002exp, Super-Kamiokande:2013ufi} and IceCube (up to 10 TeV) \cite{IceCube:2011kcp}. The study in \cite{Garcia-Cely:2017oco} translates these constraints as the lower bound on the lepton number breaking scale $v_{\phi}$ against $m_{\chi}$ above MeV. We have taken their limits from \cite{Garcia-Cely:2017oco} and imposed upon the parameter space obtained in both cases A and B.1 as shown in Figs. \ref{constraint-A} and \ref{constraint-B}. As a result, it restricts the obtained parameter space further. 
\begin{figure}[!h]
 	$
 	\includegraphics[scale=0.48]{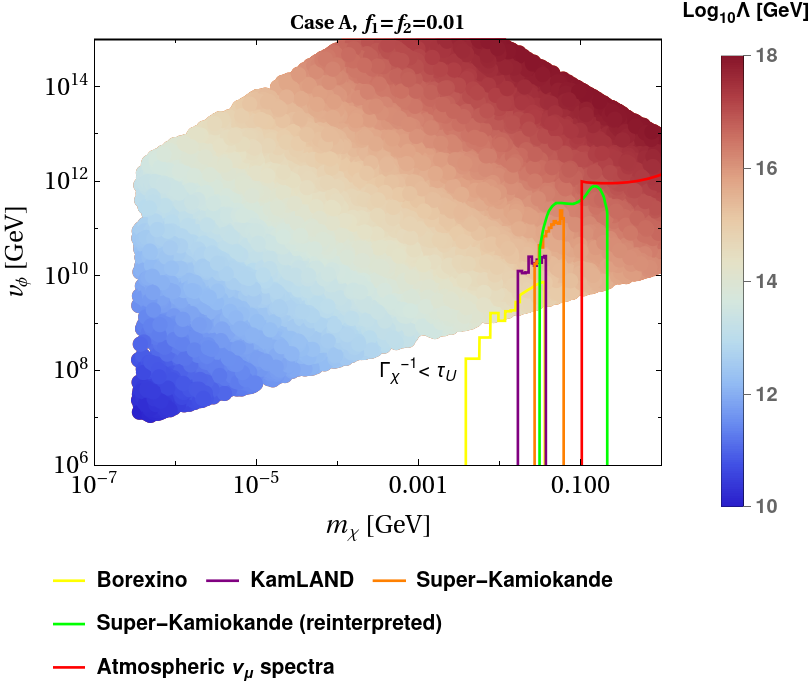}
 	$
 	\caption{$v_\phi$ (GeV) vs $m_\chi$ (GeV) parameter space for case A (when degenerate RHNs are in thermal equilibrium at $T_L$) combined with neutrino lines search experiment results. The solid lines for different experiments imply the lower bounds on $v_\phi$.}
	\label{constraint-A}
	\end{figure}

\begin{figure}[!h]
	$
	\includegraphics[scale=0.48]{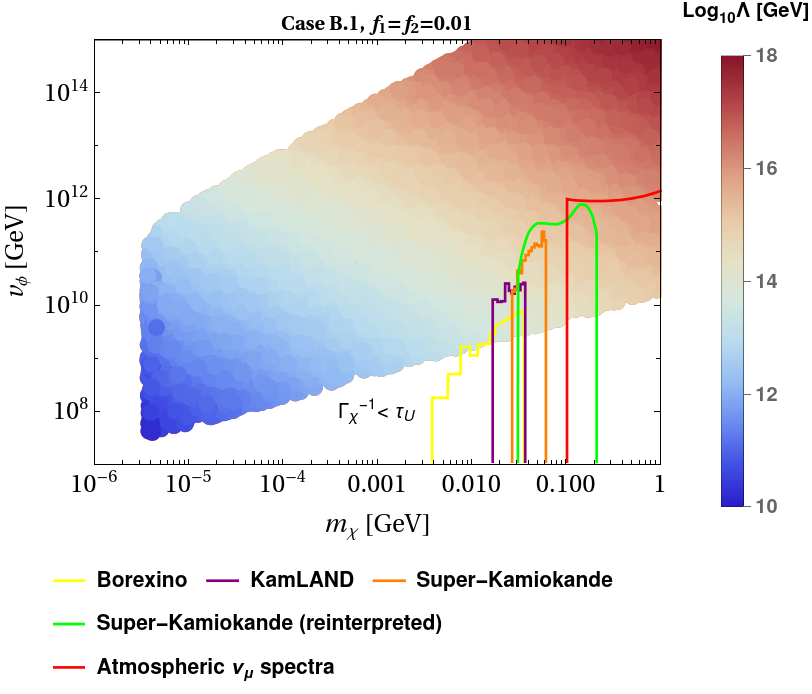}
	$
	\caption{$v_\phi$ (GeV) vs $m_\chi$ (GeV) parameter space for case B.1 (when degenerate $N_i$ are not in thermal equilibrium at $T_L$) combined with neutrino lines search experiment results. The solid lines for different experiments imply the lower bounds on $v_\phi$.}
	\label{constraint-B}
	\end{figure}
 
 \noindent  Apart from the model independent decay channel $(\chi\to\nu\nu)$, there exists another interesting decay channel of Majoron: $\chi\to \gamma\gamma$ which can be induced in two-loops \cite{Garcia-Cely:2017oco}. Following \cite{Garcia-Cely:2017oco}, a simplified expression for the associated decay width turns out to be \cite{Garcia-Cely:2017oco,Heeck:2019guh},
 \beeq
 \Gamma\left(\chi\to\gamma\gamma \right) \simeq \frac{\alpha^2 ({\rm{tr}}{\mathcal{K}})^2}{4096\pi^7}\frac{m_\chi^3}{v^2}\left| \sum_{f}N_c^f T_3^f Q_f^2 g\left(\frac{m_\chi^2}{4m_f^2} \right)\right|^2,
 \label{gamma-line}
 \eeq
 where $f$ denotes all the SM fermions having color multiplicity as $N_c^f=3 (1)$ for quarks (leptons), $T_3^f$ is the isospin, $Q$ is the corresponding electric charge in units of $e=\sqrt{4\pi \alpha}$. The loop function is given by
 \beeq
 g(x)=-\frac{1}{4x}\left(\textrm{log}|1-2x+2\sqrt{x(x-1)}| \right)^2,
 \nonumber
 \eeq 
 having $x ={m_\chi^2}/{4m_f^2}$. Here ${\mathcal{K}}$ is a model-dependent parameter given by ${\mathcal{K}}=m_D m_D ^\dagger/\left(vv_\phi\right)$, a dimensionless hermitian coupling matrix. Such $\gamma$ emission lines 
 can be probed by many experiments such as INTEGRAL $\gamma$ observatory \cite{Boyarsky:2007ge}, COMPTEL/EGRET \cite{Yuksel:2007dr}, Fermi-LAT \cite{Fermi-LAT:2015kyq}. In addition, the diffuse X-ray background observed by HEAO \cite{Boyarsky:2005us} was looking for the emission line over $3-48 $ keV range, while other X-ray telescopes like $chandra$ \cite{Bazzocchi:2008fh,Lattanzi:2013uza} and XMM \cite{Boyarsky:2007ay} cover the energy range $0.3-12$ keV. Similarly, the $\gamma$-ray line emission of $20$ keV- $7$ MeV can be covered by INTEGRAL SPI observations \cite{Boyarsky:2007ge}. Energies above $7$ MeV is covered by a combination of COMPTEL and EGRET $\gamma$-ray telescopes. Fermi 
 $\gamma$-ray telescope \cite{Fermi-LAT:2015kyq} recently searches the emission line within $7-200$ GeV energy range. 
 
  \begin{figure}[]
 	\includegraphics[scale=0.5]{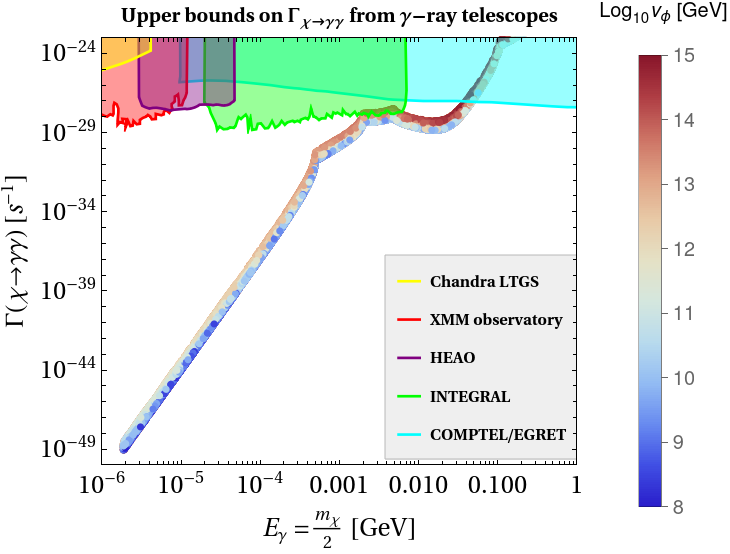}\\ 	
 \vspace{1cm}
	\includegraphics[scale=0.5]{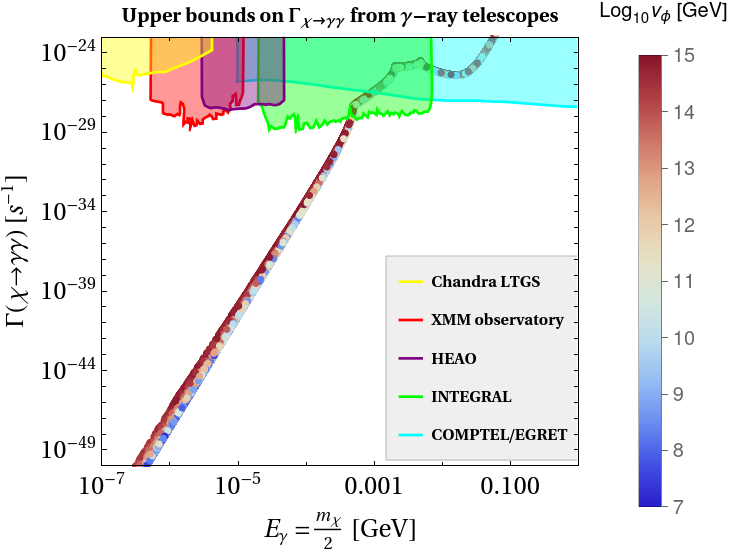}
	\caption{$\Gamma_{\chi\to\gamma\gamma}$ vs $E_\gamma=\frac{m_\chi}{2}$ with $v_\phi$ in side colourbar. \textit{Upper panel}: case B.1 $(f_1=f_2=0.01,~\theta_R=\frac{\pi}{4})$ and \textit{Lower panel}: case B.2 $(f_1=009,~f_2=0.1,~\theta_R=2.5+1.463i)$. The line emission constraints (from different $\gamma$-ray observations) are shown along with the model prediction.}
	\label{decay-gamma}
\end{figure}

 We first evaluate the ${\mathcal{K}}$-matrix for each $m_{\chi}$ from the allowed parameter space (using $v_{\phi}$ and neutrino Yukawa coupling $y^\nu$ via Eq. \ref{CI}) of Figs. \ref{PS1} and \ref{PS2}. After plugging this $\mathcal{K}$ into Eq. \ref{gamma-line}, respective $\Gamma (\chi \to \gamma\gamma)$ is plotted against $E_{\gamma} = m_\chi/2$ in top (for case B.1) and bottom (case B.2) panels of Fig. \ref{decay-gamma} for the entire allowed parameter space obtained in our proposal where the limits from different X-ray and $\gamma$-ray observations are also shown. The differently coloured shaded patches on the upper side of each plot indicate the excluded regions for different experiments, the description of which is given in the inset of each figure. It can be seen that the Majoron mass below MeV can easily evade these bounds making the cosmological bound most effective in the sub-MeV mass regime. Whereas, relatively larger values of $v_{\phi}$ appear to 
 be somewhat constrained for $m_{\chi} \sim 3.5-13$ MeV mass for case B.1. Alongside, Majoron masses beyond 100 MeV, in this case, fall within the sensitivities of the observations by COMPTEL and EGRET. For case B.2, it can be seen from Fig. \ref{decay-gamma} that the entire $m_\chi\gtrsim 0.8$ MeV region turns out to be within the exclusion limits of $X$-ray and $\gamma$-ray emission line bounds for this particular choice of $\theta_R$. However, this conclusion is not very robust as it can be altered for a different choice of $\theta_R$, and hence model dependent. This stems from the fact that the decay width $\Gamma (\chi \to \gamma\gamma)$ depends on the $\mathcal{K}$ matrix which is governed by the choice of complex 
$\theta_R$. In case of real $\theta_R$ however, the $\mathcal{K}$ remains independent of $\theta_R$. For this reason, we do not incorporate a separate figure for case A (with real $\theta_R$) which would result in a similar plot as the top panel of Fig. \ref{decay-gamma}.

It is interesting to note that the lighter side of the Majoron dark matter in our analysis, in particular 1-10 keV mass region, apparently coincides with the conventional warm dark matter (WDM) mass regime. Such keV scale WDM experiences a stringent constraint from observations of Lyman-$\alpha$ forest \cite{Viel:2013fqw,Narayanan:2000tp,Viel:2005qj,Baur:2015jsy,Irsic:2017ixq,Palanque-Delabrouille:2019iyz,Garzilli:2019qki}, thereby imposing a limit:
$m_{{WDM}}\gtrsim (1.9-5.3)$ keV at $95\%$ CL with the inherent assumption that the WDM maintains a thermal distribution. However, the Majoron here is a FIMP type DM which never attains thermal equilibrium. Hence, the above mentioned bound requires modification \cite{Heeck:2017xbu,Bae:2017dpt,Kamada:2019kpe} before applying to our case.

To have a mapping between the above constraint on conventional thermal $m_{WDM}$ onto a modified one for FIMP type DM mass $m_{FDM}$, following \cite{Bae:2017dpt,Kamada:2019kpe}, 
we define a quantity called $\it{warmness}$ associated to the DM (stands for either WDM or FDM) as $\omega_{DM}  =\tilde{\omega}_{DM} \frac{T_{DM}}{m_{DM}}$, where 
\beeq
\tilde{\omega}_{DM}^2=\frac{\int d^3 q q^2 f_{DM}(q)}{\int d^3 q f_{DM}(q)},
\label{omega-tilde}
\eeq
and $T_{DM}$ represents the temperature of the DM which is related to the temperature of the thermal bath $(T)$ via
\beeq
T_{DM}=\left(\frac{g_\star^\mathcal{S}(T)}{g_\star^\mathcal{S}(T_{dec})}\right)^{1/3}T=\mathcal{A} T. 
\label{T-chi}
\eeq
Here $T_{dec}$ corresponds to a temperature when the DM production becomes most efficient.
In Eq. \ref{omega-tilde}, the physical momentum $p$ is scaled 
as $q= p/T_{DM}$, and $f_{DM}$ corresponds to the phase space distribution of the respective DM. 

For a feebly interacting DM like Majoron as $\chi$ in our case, the phase space distribution can be obtained \cite{Kamada:2019kpe} considering the main production process: $NN\to\chi\chi$ as 
\beeq
f_\chi (q) 
\simeq \frac{1}{\mathcal{H}(T_{dec})}\int_{x_i}^{x_f} dx  \frac{x}{E_\chi} C[f_\chi],
\label{phase-space}
\eeq	
where, $x=M_1/T$, $\mathcal{H}(T_{dec})$ is the Hubble expansion rate at $T_{dec} \sim M_1$ (beyond which it freezes in) and 
$C[f_\chi]$ denotes the collision term corresponding to $\chi$ production channel:
$NN\to\chi\chi$. Using Eq. \ref{effective-int}, the simplified expression of $C[f_\chi]$ for such process can be written as \cite{Ballesteros:2020adh},
\beeq
C[f_\chi]\simeq \frac{4 g_N^2 g_\chi \Gamma(2) T^5}{(2\pi)^2\Lambda^2 p^2}\left(\frac{p}{T} \right)^2 e^{-p/T},
\label{collision}
\eeq
where $g_N=2$ and $g_\chi=1$ are the spin degrees of freedom of RHN and $\chi$ respectively.
Employing Eq. \ref{collision} into Eq. \ref{phase-space} and integrating it within the limits between $x_i=M_1/v_\phi$ (we consider the production of $\chi$ initiates at $T=T_L\sim v_\phi$) and $x_f=1$ (production of $\chi$ gets ceased at $T \sim M_1$), the phase-space distribution of Majoron can be expressed, upto a proportionality factor, as,
\beeq
f_\chi (q) \propto\int_{x_i}^{x_f}dx \frac{1}{q x \mathcal{A}(x)} e^{-q\mathcal{A}(x)}. 
\label{phase-space2}
\eeq 
With this phase-space distribution for $\chi$, we find $\tilde{\omega}_{\chi} \simeq 2.45$ whereas for a thermal scalar WDM, $\tilde{\omega}_{WDM}$ is found to be 3.22. 

As mentioned earlier, finally one aims to convert the existing lower bound on $m_{WDM}$ onto $m_{FDM}$ ($i.e. ~m_{\chi}$ here) and that is done by equating the warmness ($\omega$) 
of thermal WDM to that of $\chi$ (non-thermal Majoron) \cite{Bae:2017dpt,Kamada:2019kpe}. Using the relation between $\omega$ and $\tilde{\omega}$, such a consideration leads to 
\beeq
m_\chi=m_{WDM}\left(\frac{\tilde{\omega}_\chi}{\tilde{\omega}_{{WDM}}} \right)\frac{T_{\chi}}{T_{WDM}}. 
\label{mchi-bound}
\eeq	
While the temperature of Majoron $(T_\chi)$ can be obtained from Eq. \ref{T-chi} with $T_{dec} \sim M_1$, the temperature of the thermal WDM $(T_{{WDM}})$ 
is to be extracted from the relic satisfaction of WDM via,
\beeq
 \Omega_{{WDM}}h^2 =\frac{g_{WDM}}{2} \left(\frac{m_{{WDM}}}{94~\textrm{eV}} \right)\left(\frac{T_{{WDM}}}{T_\nu} \right)^3,
 \eeq
 where $g_{WDM}=1$ (for scalar) and $\Omega_{{WDM}}h^2 = 0.12$. 
 Using Eq. \ref{mchi-bound}, the thermal WDM mass $(m_{{WDM}})$ involved in  
the Lyman-$\alpha$ bound can be translated into $m_{\chi}$ (FIMP type DM produced via $NN\rightarrow \chi \chi$ process) as 
\beeq
m_\chi\simeq 5.3~ \textrm{keV} \left(\frac{m_{{WDM}}}{3~\textrm{keV}}\right)^{4/3} 
\left(\frac{106.75}{g_\star^\mathcal{S}(T_{dec})}\right)^{1/3}. 
\label{mchi-bound2}
\eeq
 Here $g_\star^\mathcal{S}(T_{dec})= g_\star^\mathcal{S}(M_1)\sim 106.75$ as $M_1>>T_{EW}$. Using $m_{{WDM}} = 3$ keV as the a reference value, we therefore obtain an approximate lower bound on Majoron mass from Lyman-$\alpha$ observation as, $m_\chi \geq 5.32$ keV, specific to our case.
\section{conclusion}

In this work, we relooked into the original Majoron model, the existence of which is due to the spontaneous breaking of global lepton number symmetry. Its feeble coupling with the SM fields renders itself as a natural candidate for FIMP-like dark matter. Earlier studies showed that though it is plausible to have Majorons as FIMP-like dark matter, the relic satisfaction remains a challenge and Majoron mass can either be 2.7 MeV or heavy $\sim$ TeV. In an effort to explore the possibility of having Majorons as a viable FIMP type DM candidate in a broader range (and below GeV) so as to make it interesting from the observational point of view, we include a higher order explicit lepton number breaking term in the Lagrangian involving RHNs and the spontaneous symmetry breaking scalar field. While such an interaction presents itself as a natural soft symmetry-breaking term (due to the suppression by the cut-off scale), it introduces new feeble interaction for Majorons, thereby introducing new production channel for Majorons via annihilation of the RHNs. As a result, a viable parameter space of freeze-in type of Majoron dark matter follows in the range $\mathcal{O}$(keV - GeV).
 
It is perhaps pertinent here to compare our results with other works which include such higher order Majorons-RHNs interactions. In \cite{Frigerio:2011in}, such interaction arised effectively from $f \Phi NN$ term (also responsible for RHN mass $M_N = f\langle \Phi \rangle$) via the $t$-channel RHN mediated diagram and simultaneously a contact-interaction results while using non-linear representation of pNGB field.  The associated cross-section for 
the Majoron production via $NN \rightarrow \chi \chi$ turned out to be proportional to $f^4 /M_N^2$ whereas in our case, we have an explicit LNV interaction $NN \chi^2$ which is independent of $f$ or $M_N$, thereby extending the freedom in satisfying the relic density constraint. Hence the `two Majorons and two sterile neutrinos' contribution exercised in \cite{Frigerio:2011in}, though present here too, becomes subdominant (we fix $f = 0.01$) compared to the contribution of our explicit LNV dimension 5 operator. Additionally, \cite{Frigerio:2011in} has two types of Majoron production channels: (i) $h \rightarrow \chi \chi$ followed from an explicit tree level LNV term $\lambda \chi^2 H^\dagger H$ which is absent in our analysis (as from the very beginning we consider the breaking of LN only at higher order). There the parameter $\lambda$ is also linked with $f \Phi NN$ coupling via logarithmic divergent neutrino loop. This channel turned out to be important in limiting the Majoron mass from above: $m_{\chi} \lesssim 0.0015$ GeV in \cite{Frigerio:2011in}. (ii) On the other hand, in \cite{Frigerio:2011in}, the Majoron mass had a lower limit $m_{\chi} \gtrsim 10^{-7}$ GeV where the Majoron production is dominated by $NN \rightarrow \chi \chi$ process. Along a similar line, the work of \cite{Shakya:2018qzg} also considered the $NN \rightarrow \chi \chi$ process which proceeds via $t$-channel exchange of heavy RHNs. The paper however focussed on two possible candidates of dark matter: lightest RHN $N_1$ and Majoron $\chi$. With $m_{\chi} < M_N$, the case is similar to that of \cite{Frigerio:2011in}. 

In a nutshell, the presence of a separate dimension-5 operator in our work relaxes the tension among the parameters which results an extended parameter space for $m_{\chi}$. 
Our analysis shows that an adequate region of parameter space consistent with Majoron as dark matter satisfying correct relic abundance in the keV-GeV mass range can be obtained. While this entire mass regime is important in the context of supernova cooling, the above-MeV region is also fascinating to be probed by monochromatic neutrino lines. Even though the bounds from such experiments turn out to be more stringent than the naively considered lifetime of the Universe, the parameter space undergoes a rich dark matter phenomenology and finally provides compelling evidences in the form of gamma-ray lines that can be probed by ongoing and future experiments indicating some tantalizing connection between dark matter and neutrino physics.

\begin{acknowledgements}
The work of AS is supported by the grants CRG/2021/005080 and MTR/2021/000774 from SERB, Govt. of India. SKM would like to thank Arghyajit Datta for various useful discussions.
\end{acknowledgements}

\bibliography{ref.bib}

\end{document}